\documentclass[fleqn,12pt,twoside]{article}
\usepackage[headings]{espcrc1}

\readRCS
$Id: espcrc1.tex,v 1.2 2004/02/24 11:22:11 spepping Exp $
\usepackage{graphicx}
\usepackage[figuresright]{rotating}

\newcommand{\AmS}{{\protect\the\textfont2
  A\kern-.1667em\lower.5ex\hbox{M}\kern-.125emS}}

\title{Experimental Approach to Stellar Reactions with RI Beams \\
- Overview of Experiments on Hydrogen Burning -}

\author{S. Kubono\address[CNS]{Center for Nuclear Study (CNS), University of Tokyo,
        Wako Branch at RIKEN, \\
        Hirosawa 2-1, Wako, Saitama 351-0198, Japan}%
\thanks{Invited talk presented at the IX International Conference on Nucleus Nucleus Collisions, Rio de Janeiro, August 28 – September 1, 2006, and to be published in Nucl. Phys. A.
                }
}

\runtitle{ Experimental Approach to Stellar Reactions with RI Beams}
\runauthor{S. Kubono}

\begin{document}

\maketitle

\begin{abstract}
After a short review on resent developments achieved in astrophysics in the past years since last NN conference, experimental efforts in nuclear astrophysics primarily with RI beams were revisited, especially on the works relevant to neutron-deficient nuclei, the other half of the nuclear chart reviewed by Rehm in this conference.  A new interesting recognition discussed in the past years is the important role of explosive hydrogen burning process in the very early stage of type II supernovae. A new broadening research field related to the first generation stars both from observations as well as from nuclear astrophysics was also discussed.

\end{abstract}

\section{PROGRESS IN ASTROPHYSICS}

As is well known, nuclear physics plays a key role in the evolution of the universe as well as in various stellar phenomena in terms of synthesis of elements and energy generation.  Specifically, the investigation of the nuclear phenomena under high temperatures and high densities in the universe has attracted many nuclear physicists because nuclear physics can be a key element for understanding the astronomical explosions such as supernovae and gamma-ray burst\cite{ku1}.  

Astronomical observations have broadened related fields of science for the universe, especially since the observation of Supernova 1987A.  Nuclear Astrophysics is one of the basic fields developed rapidly since then for understanding astronomical explosive phenomena.  Recent observation of nuclear gamma rays by satellites\cite{di} and successful identification of isotopic ratios of some elements by high-resolution optical telescopes\cite{ao} really provide indispensable information on the critical nuclear reactions relevant to the phenomena and thus on the amount of isotopes produced in the explosive phenomena.  This information is a stringent test and help for establishing the model. 

In addition in astronomy, continuous and extensive efforts have been made on research for the first generation stars.  One of the recent striking reports is the observation of extremely metal poor stars, that have irons only of a part of 3x10$^{5}$ of the solar abundance, but overabundant for the light elements such as C, N, and O.  Further, one observation of such stars reports a significant amount of Sr and some others among heavy elements\cite{ao2} in addition. This will be an important clue for understanding the mechanism of evolution of such early stars. Clearly, it demonstrates that the role of nuclear physics becomes more important in astrophysics. Evolution models for the early stars really need precise nuclear physics information. 

Another important facet recognized in the past years is that a proton-rich environment play a key role in the very early stage of type II supernovae because of the neutrino processes.  This process is called $\nu$p-process\cite{fr,wa}.  The nucleosynthesis flow is very much similar to the high-temperature rp-process like in X-ray bursts, but with considerable neutrons. Thus, the so-called waiting points and bottlenecks would be by-passed by the (n,p) and (n,$\gamma$) reactions at some stage.  A possible significance of the $\nu$p-process for the cosmic chemical evolution is that it could be one of the origins for the p-nuclei up to around mass 100. This new facet implies that the whole area on the nuclear chart is the target of research for understanding type II supernovae.  There is also very interesting related regime in nuclear physics, which is not studied yet, such as equation of state of nuclear matter.

We discuss the problems of the pp-chain in sec. 2, the CNO cycle and the breakout in sec. 3, the early stage of the rp-process in sec. 4, nucleosynthesis of the rp-process at the medium mass region in sec. 5, and the scope of the field for conclusion.

\section{THE pp-CHAIN and HOT pp-CHAIN}

The pp-chain is the main source for energy generation in low mass stars at the main sequence stage, where hydrogen burns under high-temperature and quiescent conditions.  The pp-chain also could play a crucial role in massive but extremely metal-poor stars.  Nuclear burning in our sun has been long investigated, both by the solar neutrino observations and laboratory studies of the nuclear reactions relevant.  The main problem of the neutrino flux deficit has been solved by finding the neutrino oscillation \cite{ah}.  However, the investigation of the solar model still needs nuclear data at very low energies with high precision for the reaction rates of the pp-chain.  The data are not available for most reactions at the energies we need.  Furthermore, the data available at higher energies have large uncertainties for the cross sections.  Among the reaction chain that leads to production of the high-energy neutrino emitter $^{8}$B, the sensitivity of each astrophysical S-factor is predicted to be \cite{bah}; 

$\Phi$($^{8}$B) = S$_{11}$$^{-2.6}$S$_{33}$$^{-0.43}$S$_{34}$$^{0.81}$S$_{17}$$^{1.0}$S$_{e-7}$$^{-1.0}$

Therefore, primarily the two reactions, $^{3}$He($^{4}$He,$\gamma$)$^{7}$Be, and $^{7}$Be(p,$\gamma$)$^{8}$B have to be determined with a high precision, like 5 \%.   

The $^{3}$He($^{4}$He,$\gamma$)$^{7}$Be reaction has been studied directly at low energies, but still the minimum experimental energy is about 150 keV, which is much higher than the Gamow energy, about 18 keV.  A recent work with a direct measurement reports a precise determination in the energy range studied before, and the result agrees well with the previous ones with better experimental uncertainties.  Their value is S$_{34}$(0)=0.53 (2)(1) keVb, where the parentheses indicate the systematic and statistical uncertainties, respectively \cite{na}.  
More interesting and rigorous challenge for experiments is to go underground laboratories to reduce the cosmic ray background with the direct method.  Such a project is underway at the Gran Sasso Laboratory in Italy\cite{bem}.  This is one of the frontiers in experimental nuclear astrophysics.  One needs to measure the cross sections at lower energies towards the Gamow energy regions, where one has extremely small cross sections with high relative background.

The other crucial reaction for the solar model is $^{7}$Be(p,$\gamma$)$^{8}$B. There are many extensive works reported, but they deviate considerably each other yet.  Now we need to clarify the problems.  One needs to investigate carefully both from nuclear structure point of view as well as from reaction dynamics point of view.  If, for instance, there is a broad s-wave resonance at around 3.5 MeV in $^{8}$B, as suggested in ref. \cite{rog}, it might have a chance to influence on the extrapolation of S$_{17}$ to the Gamow energy region.  An interesting experiment was recently performed at the low-energy in-flight RI beam separator CRIB of the University of Tokyo\cite{yan,ku6}.  The 2$^{-}$ s-wave resonance has been clearly identified, and in addition some new resonances have been discovered at higher excitation energies  by a resonant scattering of protons from $^{7}$Be+p using a thick target method\cite{yam}.  

It is also important to investigate from reaction mechanism point of view for this problem.  The Coulomb dissociation method has been utilized to study efficiently the low energy (p,$\gamma$) reaction cross sections, but the S$_{17}$ values derived by a simple DWBA analysis tend to show a little smaller values than those obtained by other methods. This method has potentially some problems in deriving precisely the cross sections of interest.  Since high energy heavy ions are used, the nuclear contribution and higher order processes are inevitably involved.   Recently, a detailed analysis of the Coulomb dissociation process including coupling to the continuum states has been made \cite{es,og}, giving an S$_{17}$ value slightly larger, and consistent with the values obtained by other methods.

Hydrogen burning in the mass region below mass 12 has a potential importance for nucleosynthesis in metal poor stars such as very early stars, which had mostly H and He from the Big Bang nucleosynthesis.  Under such environment, it is usually believed that the main path to produce heavy elements like C and O is primarily the triple $\alpha$ process. However, under very metal poor condition, C and O could be produced by the reaction chain that goes through the proton-rich side bypassing the A=8 mass gap.  This postulated pathway is called the hot pp-chain \cite{wi}. This proton rich region is not much investigated yet.  The important path ways are considered to be $^{7}$Be(p,$\gamma$)$^{8}$B(p,$\gamma$)$^{9}$C ($\alpha$,p)$^{12}$N($\beta^+$)$^{12}$C and 
$^{7}$Be($\alpha$,$\gamma$)$^{11}$C(p,$\gamma$)$^{12}$B($\beta^+$)$^{12}$C.  Here, $^{11}$C as well as $^{9}$C sits on the branching point for breakout from the hot pp-chain to the CNO cycle.  Although there are some experiments of $^{11}$C(p,$\gamma$)$^{12}$B made by the Coulomb dissociation method \cite{min,le} and the ANC method\cite{tr}, there are still large uncertainties in the reaction rate.  Thus, it is still of great interest to study directly the $^{11}$C(p,$\gamma$)$^{12}$B reaction. This reaction will be investigated in near future at extensive low-energy RI beam facilities like TRIUMF.  As was discussed in sec. 1, these data are really needed to investigate the evolution of early stars and the very early epoch of type II supernovae.

\section{CNO-CYCLE AND THE BREAKOUT}

In this mass region under hydrogen-rich environment, we have typically the CNO cycle and the extended CNO cycle.  These processes play a basic role in evolution of stars. There are still many reactions among the cycle to be investigated further to clarify the burning process.  On the other hand, there are many observational data of elemental abundance relevant, and thus this mass region has been extensively investigated in the past years using both stable and unstable nuclear beams. In addition, the breakout process from the hot-CNO cycle is also an interesting subject for studying the nature of explosive stellar events. 

There are many experimental works reported meantime for this mass region. I specifically touch on two stellar reactions, i.e., 
  $^{14}$N(p,$\gamma$)$^{15}$O and $^{18}$F(p,$\alpha$)$^{15}$O \cite{de,ko}.  The former reaction is the slowest process among the CNO cycle, and thus it has a decisive role for stellar evolution, such as for the age of globular clusters.  The latter reaction is related to the flux of positron-electron annihilation gamma rays emitted in the occasion of novae, because $^{18}$F is considered to be one of the main contributors for the annihilation gamma rays in the events.  The (p,$\alpha$) reaction destroys the nucleus $^{18}$F.  Now, the INTEGRAL project is reporting an observation of such annihilation gamma rays \cite{di}.  Thus, a quantitative estimate of the annihilation gammas is really awaited for from astronomy.  

There are several works reported on the $^{14}$N(p,$\gamma$)$^{15}$O reaction.  Previously, Schroder reported an extensive work on this reaction\cite{sch}.  Recently, Angulo\cite{an} reanalyzed the data and found that the summing effects was not properly corrected in the gamma ray measurements, finding much smaller reaction rates for the component to the ground state transition.  Two experiments recently have confirmed this effect\cite{fo1,run}, and the measurements were further extended to lower energy region toward the Gamow peak energy.  In fact, the experiment of LUNA project at Gran Sasso has just reached the high-energy end of the Gamow energy window \cite{fo2}. The Gamow window will be covered in near future in the LUNA project.  This is a beautiful accomplishment for the low energy frontier in nuclear astrophysics.

Another interesting, and crucial subject for this mass region is the breakout process from the CNO Cycle.  There are two breakout processes suggested; one starting from $^{14}$O and the other one from $^{15}$O.  Under a high temperature and high density condition just before the breakout, the most abundant nuclides are $^{14}$O and $^{15}$O together with abundant $\alpha$ particles, and thus the $\alpha$ induced reactions on $^{14}$O and $^{15}$O can be the ignition reactions. Namely, the reaction chain of $^{15}$O($\alpha$,$\gamma$)$^{19}$Ne(p,$\gamma$)$^{20}$Ne is considered to be the key process under novae condition, but the current estimate seems unlikely.  However, none of the reactions in the process is known well yet.  The other process $^{14}$O($\alpha$,p)$^{17}$F(p,$\gamma$)$^{18}$Ne($\alpha$,p)$^{21}$Na is considered to be the key process under high temperature rp-process like in X-ray bursts.  Again, no reaction in the chain is well understood at the stellar temperatures. 

 The Notre Dame Group has made a great progress in the past years for the reaction $^{15}$O($\alpha$,$\gamma$)$^{19}$Ne.  This reaction is considered to be one of the toughest reaction to study directly with the RI beam of $^{15}$O because it would require at least 10$^{10}$ pps of $^{15}$O.  Instead, they utilized an indirect method, namely investigated the resonance parameters of the possible key resonance, the 4.033 MeV 3/2$^{+}$ state.  Since this state is close to the $\alpha$ threshold, the $\alpha$ width is considered to be much smaller than the $\gamma$ width.  Thus, the reaction rate may be primarily determined by the $\alpha$ width.  They determined the width of the resonance by measuring the life time of the resonance using the Doppler shift attenuation method.  The result is $\Gamma_{\gamma}$ = 51 meV \cite{go}.  The lower limit of the $\alpha$ width was determined in a separate run by measuring the decay $\alpha$ particles from the resonance\cite{ta}.  Further experiment will provide the reaction rate of the resonance term in $^{15}$O($\alpha$,$\gamma$).  

The other progress was made on the $^{14}$O($\alpha$,p)$^{17}$F reaction\cite{no} using the low energy RI beam of $^{14}$O at CRIB\cite{yan} of the University of Tokyo.  The reaction cross section was measured by a thick target method of a cooled He gas target.  Here, the He gas target was cooled down to about 30K that reduced the target length roughly by a factor of ten, which made the experiment much easier.  The ejectiles, protons were measured by a counter telescope of Si detectors set at zero degree.  Since the thick target method was applied \cite{ku2}, the proton energy spectrum is, in principle, the same as the excitation function of the ($\alpha$,p) reaction. The several peaks observed correspond well to the resonance energies known for the excited states in $^{18}$Ne.  The important result is that the cross sections to the levels around  Ex = 6.2 MeV in $^{18}$Ne were determined well and consistent with the previous predictions.  The other interesting observation is that the ($\alpha$,p) reactions to the first excited state in $^{17}$F was observed with an intensity of about a half of that to the ground state, implying that the reaction rate to be increased by about 50 \%.  It should be of great interest to investigate further, because the measurement was made only at zero degree and has some uncertainties of other excited state contributions.  There are some other efforts on this reaction with $^{14}$O beams.
  
There are not much progresses reported for other reactions among the breakout process, such as $^{18}$Ne($\alpha$,p)$^{21}$Na and $^{19}$Ne(p,$\gamma$)$^{20}$Na since the last NN conference.  The $^{18}$Ne($\alpha$,p)$^{21}$Na reaction was investigated directly some time ago for the high temperature region \cite{gr}.  The latter reaction is a long-standing problem for the breakout of the hot-CNO cycle \cite{ku3}, but it is still not well understood.  So far, the estimates suggest that this pathway is less feasible for the breakout under the nova conditions.  These reactions also are an important subject to be investigated further.

\section{EARLY STAGE OF THE rp-PROCESS}

In this section, we will discuss research activities in the mass region of A = 20 $\sim$ 40, where the early stage of the rp-process takes place, and astronomically interesting nuclei $^{26}$Al and $^{22}$Na are involved.  The nuclear gamma rays from the decay of $^{22}$Na is a keen object of the nuclear gamma ray observation by the INTEGRAL project\cite{di}, although it has not been detected clearly yet.

There are many works investigating nucleosynthesis in explosive hydrogen burning in the past years in this mass region both by direct and indirect methods.  An interesting observation is that a variety of experimental methods have been used for spectroscopy to deduce reliable reaction rates.
They include a Penning trap measurement for precise mass determination, Gamma Sphere for full spectroscopy of the threshold states, particle spectroscopy, etc. 

A series of experiments were performed previously \cite{ku4} using indirect methods for a possible pathway just after the breakout, $^{19}$Ne(p,$\gamma$)$^{20}$Na(p,$\gamma$)$^{21}$Mg($\beta^{+}$)$^{21}$Na(p,$\gamma$) \\ 
$^{22}$Mg($\beta^{+}$)$^{22}$Na(p,$\gamma$)$^{23}$Mg(p,$\gamma$)$^{24}$Al(p,$\gamma$)$^{25}$Si.  Many new resonances just above the proton thresholds were discovered by using particle transfer reactions such as ($^{3}$He,$^{6}$He) \cite{ku5}.  However, resonance parameters are needed to determine the reaction rate of their contributions. The resonance parameters can be better investigated directly using the proton-rich unstable nuclear beams.  Resonant proton elastic scattering provides information on the incident channel of the proton capture reactions of the rp-process. A series of such experimental studies are in progress using low-energy proton-rich RI beams at CRIB.

A great advance in experimental nuclear astrophysics was reported from TRIUMF \cite{bi}, where one of the most advanced low-energy RI beam facility, based on ISOL system, has become operational and succeeded in performing a direct measurement of the stellar reaction $^{21}$Na(p,$\gamma$)$^{22}$Mg using the RI beam of $^{21}$Na accelerated. They measured directly the cross sections of the proton capture reaction at the stellar energies of interest with a recoil mass separator DRAGON together with an array of gamma-ray detectors.  The experiment was guided by the precise excitation energies determined beforehand by the $^{24}$Mg(p,t)$^{22}$Mg reaction at CNS, University of Tokyo\cite{bat}.  This is a good example that the indirect method is very useful and complementary to the direct method.  The resonance strength of the resonance at 206 keV, which dominates the reaction rate in the nova condition, has been successfully determined experimentally for the first time.  This is a beautiful, extensive work that directly determined for the first time the reaction rate.  The TRIUMF project also succeeded in determining the reaction rate of the $^{26}$Al(p,$\gamma$)$^{27}$Si stellar reaction \cite{rui}.  These successes were brought by the development of the first extensive RI beam facility in the world, where a high intensity low-energy RI beams, produced by a high-intensity 500 MeV proton beam on a thick production target, were accelerated by a linear accelerator to the astrophysical energies.     

On this specific stellar reaction of $^{21}$Na(p,$\gamma$)$^{22}$Mg, other powerful spectroscopic tools well known for nuclear spectroscpy have been utilized for indirect methods.  After the success of the TRIUMF experiment, a discrepancy was identified for the mass of $^{22}$Mg.  Later, the mass was precisely measured\cite{mu} using an ion trap, and the properties of the resonances were examined experimentally using the Gamma Sphere\cite{se}.  These have helped to establish well the reaction rate of $^{21}$Na(p,$\gamma$)$^{22}$Mg.

\section{WAITING POINT AND BOTTLENECK OF THE rp-PROCESS}

Under extremely high-temperature and high density conditions of hydrogen, the rp-process is expected to run up to the mass region of around A = 100. This is a typical process in X-Ray burst \cite{sh}.  Experimental efforts in this mass region just have started in the last several years, and thus most part of the rp-process is not well known yet.  It is still the early stage of the work as compared to the low mass regions discussed above.  So, the pathway is of course one of the key issues for the nucleosynthesis.  The masses and half-lives roughly determine the pathway. Of course, the detail of nuclear properties eventually decides each reaction rate, and consequently the pathway exactly.  These discussions apply critically to the possible nuclei at the waiting points and bottlenecks. 

The masses of the possible waiting point nuclei were precisely determined using Penning traps; the mass of $^{68}$Se at ANL\cite{cl} and $^{72}$Kr at ISOLDE\cite{ro}.  Similarly, the masses of $^{41}$Ti, $^{44}$V, $^{45}$Cr, and $^{48}$Mn were also measured successfully by ESR at GSI \cite{st}.  The structure of the proton threshold states in $^{69}$Se and $^{70}$Br need to be investigated whether the nucleosynthesis flow really favors to go through the bottleneck nuclei.

Another interesting challenge reported is to clarify experimentally the endpoint of the rp-process.  Shatz predicted it to be around mass $^{107}$Te\cite{sh}.  A detailed spectroscopic study is in progress at Michigan State University (MSU).  The MSU group also reported a new method to investigate the resonant states just above the proton threshold of the rp-process nuclei to deduce astrophysical (p, $\gamma$) reaction rates \cite{cle}. A Detailed discussion on the MSU activities will be found in the contribution by H. Schatz in this proceeding.

The $\nu$p-process would run through the same mass region at the very early stage of type II supernovae, but the time duration at the waiting point will be very much reduced by (n,p) and (n,$\gamma$) reactions.  These could contribute to the p-nucleus production in the $\nu$p-process\cite{fr}.

\section{SCOPE}

Although I discussed mostly the stellar reactions that involve neutron-deficient nuclei, there are many other important and interesting problems in the nucleosynthesis of the nuclei near the line of stability and the nuclei of neutron-rich side, which are fully reviewed by Rehm in this meeting.  Since I confined my talk to the neutron-deficient side of the nuclear chart, I did not touch on the problems of s-process, which is also important for heavy element synthesis and consequently for explosive nucleosynthesis because one needs to untangle the problem in the region. There is also other crucial nuclear physics research needed for nuclear astrophysics.  The very early epoch of type II supernovae has an extremely high flux of neutrinos that play a decisive role for the explosion as well as for nucleosynthesis.  These neutrino-nucleus process needs to be investigated.  Another crucial physics parameter is the Equation of State of nuclear matter. This will have a decisive role for the explosion mechanism of type II supernova models.  It is partly discussed by Lynch\cite{fr} in this meeting.  These subjects should be eventually investigated again in the unstable nuclear region, and they are the important themes in the third generation RIB facilities such as RI Beam Factory at RIKEN, which starts operation very soon, and FAIR Project at GSI.  There are many other RI beam facilities coming up in the world that will provide a variety of opportunities for nuclear astrophysics.  As I demonstrated the case of CRIB here, even a small accelerator facility, one can obtain quite high intense RI beams that enable one to make important contribution for this field.

\end{document}